\documentclass[11pt]{article}
\usepackage{theorem}
\usepackage{latexsym}
\usepackage{amsmath}

\usepackage[dvips]{graphicx}
\usepackage{graphicx}

\newcommand{\newc}{\newcommand}

\newc{\be}{\begin{equation}}
\newc{\ee}{\end{equation}}
\newc{\bea}{\begin{eqnarray}}
\newc{\eea}{\end{eqnarray}}
\newc{\beas}{\begin{eqnarray*}}
\newc{\eeas}{\end{eqnarray*}}

\newc{\pardt}{\partial_{t}}
\newc{\pardxi}{\partial_{i}}
\newc{\pardts}{\partial_{t^{*}}}
\newc{\pardxis}{\partial_{i^{*}}}
\newc{\pardxj}{\partial_{j}}
\newc{\pardxk}{\partial_{k}}
\newc{\pard}{\partial}

\newc{\s }{\overline}
\newc{\sect}{\section}
\newc{\subs}{\subsection}

\newc{\defi}{\definition}
\newc{\prop}{\proposition}
\newc{\rem}{\remark}
\newc{\lem}{\lemma}
\newc{\exa}{\example}
\newc{\theo}{\theorem}
\newc{\coro}{\corollary}
\newc{\post}{\postulate}
\newc{\state}{\statement}

\begin{document}
\baselineskip0.5cm
\renewcommand {\theequation}{\thesection.\arabic{equation}}
\title{Hydrodynamic equations of anisotropic, polarized and inhomogeneous superfluid vortex tangles}

\date{ }

\author{D.~Jou$^1$, M.S.~Mongiov\`{\i}$^2$ and M. Sciacca$^2$}

\maketitle
\begin{center} {\footnotesize $^1$ Departament de F\'{\i}sica, Universitat Aut\`{o}noma de Barcelona,
08193 Bellaterra, Catalonia, Spain\\
$^2$ Dipartimento di Metodi e Modelli Matematici Universit\`a di
Palermo, 
Palermo, Italy}

PACS number(s): 67.25.dk, 47.37.+q, 67.25.dm
\end{center} \footnotetext{E-mail addresses: david.jou@uab.es
(D. Jou), mongiovi@unipa.it (M. S. Mongiov\`{\i}), msciacca@unipa.it
(M. Sciacca)}

\begin{abstract}
We include the effects of anisotropy and polarization in the
hydrodynamics of inhomogeneous vortex tangles, thus generalizing
the well known Hall-Vinen-Bekarevich-Khalatnikov equations, which
do not take them in consideration. These effects contribute to the
mutual friction force ${\bf F}_{ns}$ between normal and superfluid
components and to the vortex tension force $\rho_s{\bf T}$. These
equations are complemented by an evolution equation for the vortex
line density $L$, which takes into account these contributions.
These equations are expected to be more suitable than the usual
ones for rotating counterflows, or turbulence behind a cylinder,
or turbulence produced by a grid of parallel thin cylinders towed
across a superfluid, because in these situations polarization is
expected to play a relevant role.
\end{abstract}

\section{Introduction}

The possibility to use turbulent superfluids to explore with
relative facility turbulent flows with high Reynolds numbers is an
increasingly exciting perspective. The main attractive issue of
superfluid turbulence are quantized vortices, which are
filamentous vortices, whose core dimension is of the order of the
atomic diameter of the helium atom, of the order of 1 $\AA$, due
to the rotational of superfluid component whose circulation is
quantized. From a microscopic point of view, they are described by
specifying the whole detailed curve of each vortex line, but from
a macroscopic perspective this detailed description is lost and it
is usually reduced to simply a scalar quantity $L$, the vortex
line density, i.e., the total length of vortex lines per unit
volume.

Quantized vortices in superfluids have been mainly studied in two
typical situations: rotating superfluids and counterflow
experiments, the latter meaning the presence of a heat flux, but
with zero barycentric motion. In these situations the vortices are
modeled, respectively, as an array of parallel rectilinear
vortices or as an almost isotropic tangle. In both cases, the
mutual force between the normal component and the superfluid due
to the presence of vortex lines is well known, and the so-called
vortex line tension, due to the curvature of vortex lines, is zero
\cite{Don-book}--\cite{Vinen-JLTP128-2002}.

However, in other situations, as rotating counterflow or non
stationary Couette and Poiseuille flows, or behind a grid towed
across a quiescent superfluid, one expects a partially polarized
vortex tangle, as a compromise between the orienting effect of a
rotation or of a macroscopic velocity gradient, and the
randomizing effect of the relative velocity of normal and
superfluid components. In these cases, the mutual friction force
${\bf F}_{ns}$ between these components,  as well as the
nonvanishing tension of the vortex lines $\rho_s\bf T$, which are
fundamental ingredients of the hydrodynamics of turbulent
superfluids in the well-known Hall-Vinen-Bekarevich-Khalatnikov
(HVBK) model \cite{Hall-PRS238-1956,Beka-SPJETP13-1961}, are not
sufficiently known. Thus, the exploration of ${\bf F}_{ns}$ and
$\bf T$ for partially polarized tangles with nonvanishing average
curvature of vortex lines is an open topic.

The quantitative values of the corrections are expected to be
relevant in some steady states (as rotating counterflows, or
turbulence behind a cylinder, or turbulence produced by a grid of
parallel thin cylinders towed across a superfluid) where a
sufficiently high polarization may induce macroscopic eddies
\cite{Zhang-vanSciver}, and in many of the proposed experiments
which involve nonstationary flows, for their own sake or in view
to the application of classical measurement techniques to
superfluids, as for instance thin oscillating wires or
small-particle velocimetry. Thus a basic understanding of these
effects is essential to obtain the form of ${\bf F}_{ns}$  and
$\bf T$. This is the aim of this paper.

The structure of the paper is the following one: in Section 2 an
introduction to the usual HVBK model is made, with special
emphasis on the expression of the mutual friction force ${\bf
F}_{ns}$  and the tension $\bf T$ in the cases of rotating helium
and counterflow superfluid turbulence; in Section 3 a microscopic
expression for the friction force and vortex tension is given,
checking their limit of validity in HVBK model; in Section 4 a
more general expression of the HVBK model is given through more
exhaustive expressions for the mutual friction force ${\bf
F}_{ns}$ and tension $\bf T$. The model is completed by a
generalization of the Vinen's equation, as an evolution equation
for the vortex line density $L$. At last, in Section 5 an
application of the model to the interesting case of rotating
counterflow turbulence is made, which takes advance of the recent
study on the anisotropy of the vortex tangle.

\section{HVBK model. Evolution equations for the normal and the
superfluid components} \setcounter{equation}{0} It is known  that
if a superfluid ($^4$He and $^3$He liquid helium, Bose-Einstein
condensates, neutron stars, ...) rotates at a constant angular
velocity $\Omega$, exceeding a critical value $\Omega_c$, an
ordered array of quantized vortex lines of equal circulation
$\kappa$ parallel to the rotation axis is created
\cite{Don-book}--\cite{Vinen-JLTP128-2002}. The quantum of
vorticity is $\kappa=h/m_4$ in superfluid $^4$He, with $h$ the
Planck constant and $m_4$ the mass of  $^4$He atom; in $^3$He and
neutron stars, which are formed by fermionic particles, the
superfluidity is due to the formation of Cooper pairs, and
therefore it is {$\kappa=h/2m$}, where $m$ is the mass of an
$^3$He atom or of a neutron, respectively.

In most literature, the motion of a superfluid is modeled using
Landau's two-fluid model, which regards the fluid to be made of two
completely mixed components: the normal fluid and the superfluid,
with densities $\rho_n$ and $\rho_s$ respectively, and velocities
${\bf v}_n$  and ${\bf v}_s$ respectively, with total mass density
$\rho$ and barycentric velocity ${\bf v}$ defined by
$\rho=\rho_s+\rho_n$ and $\rho {\bf v}=\rho_s {\bf v}_s+\rho_n {\bf
v}_n$. The first component is related to thermally excited states
(phonons and rotons) that form a classical Navier-Stokes viscous
fluid. The second component is related to the quantum coherent
ground state and it is an ideal fluid, which does not experience
dissipation neither carries entropy.

If the superfluid is put in rotation with angular velocity
$\Omega$ higher than $\Omega_c$, the ordered array of parallel
quantized vortex lines is described by introducing the {line
density} $L$, defined as the average vortex line length per unit
volume, equivalent to the areal density of vortex lines, which is
proportional to the angular velocity
\cite{Don-book}--\cite{Vinen-JLTP128-2002}, namely $L= {2\Omega/
\kappa}.$

It is well known too that a disordered tangle of quantized vortex
lines is created in the so-called counterflow superfluid
turbulence \cite{Don-book}--\cite{Vinen-JLTP128-2002},
characterized by no matter flow but only heat transport, exceeding
a critical heat flux $q_c$. When the turbulence is fully
developed, the line density $L$ is proportional to the square of
the averaged counterflow velocity vector
\be\label{Vns-medio}%
{\bf
V}_{ns}= [{\bf v}_{ns}]_{av}= {1\over \Lambda} \int {\bf v}_{ns}
d\Lambda, %
\ee related to heat flux (${\bf v}_{ns}={\bf v}_n-{\bf v}_s$ being
the microscopic counterflow velocity)
\cite{Don-book}--\cite{Vinen-JLTP128-2002}, $L  \simeq \gamma_H^2
{{\bf V}_{ns}^2/\kappa^2},$ the dimensionless coefficient
$\gamma_H$ being dependent on the temperature. In
(\ref{Vns-medio}) and in the following, capital letters denote
local macroscopic velocities averaged over a small mesoscopic
volume $\Lambda$, threaded by a high density of vortex lines.

In this paper we will use two different averages which, for the
sake of clarity, we explicit here. The first one is that used in
(\ref{Vns-medio}) where the integration is made over a small
volume $\Lambda$, whereas the other one will be made on all
vortices contained in the same volume  $\Lambda$, that is
\be\label{average}
< - >={1\over L \Lambda} \int - \; d\xi,%
\ee
 where $L$ is the vortex line density in the volume  $\Lambda$
and $\xi$ is the parameter indicating the arc-length along the
vortex line, and the integral is made on all vortices contained in
$\Lambda$. The two averages are linked by the relation $[-]_{av}=L
< - >$.

Due to the smallness of the quantum of circulation, even a
relatively weak rotation  or a small counterflow velocity produces
a large density of vortex filaments. It is therefore possible to
develop a set of macroscopic hydrodynamic equations which average
over the presence of many individual vortex lines and incorporate
the macroscopic effects of the vortices in the evolution equations
for superfluid and normal fluid velocities.

As an illustration of the order of magnitude of the values of $L$
in current situations, let us mention that in counterflow in a
cylindrical channel the line length density $L$ satisfies the
relation $L^{1/2}= \gamma V_{ns}- {b\over d}$, where $d$ is the
diameter of the channel (see for example
 \cite{Martin-Tough-1983}  and  \cite{Mong-Jou-JPCM-2005});
coefficients $\gamma$ and $b$ assume different values in different
turbulent regimes. In particular in \cite{Mong-Jou-JPCM-2005} it
has been shown that, for values of the dimensionless quantity $x=
\frac{V_{ns} d}{\kappa}$ greater than 219 at $T=1.5$ K and greater
than 186 at $T=1.7$ K, the system undergoes a transition from the
turbulent TI state, with a low density of vortex lines, to the
fully developed turbulent TII state, with a high density of vortex
lines. The experimental results shows that, at the transition
TI-TII in a volume of 1 cm$^3$ there is a vortex tangle of about
total length ${\cal L}= 400$ cm.

A set of hydrodynamical equations frequently used  is the
Hall-Vinen-Bekarevich-Khalatnikov (HVBK) model
\cite{Don-book,Hall-PRS238-1956,Beka-SPJETP13-1961},
\cite{Swanson-PRL67-1991}--\cite{Henderson-TCFD18-2004}. These
equations, which were derived by a number of researchers over the
years, are those of the two-fluid model, modified to incorporate
the presence of vortices. Here, for sake of simplicity, we
consider the incompressible HVBK equations, which in an inertial
frame are written \cite{Don-book,Henderson-TCFD18-2004}:
$\nabla\cdot{\bf V}_n =0$, $\nabla\cdot{\bf V}_s =0$, and
\be\label{equa-Vn}%
\rho_n  {\partial{\bf V}_n\over\partial t}+\rho_n({\bf V}_n\cdot
\nabla){\bf V}_n = - {\rho_n\over\rho}\nabla p_n
 - \rho_s S  \nabla T + {\bf F}_{ns}+\eta\nabla^2 {\bf V}_n,%
\ee
\be\label{equa-Vs}%
\rho_s  {\partial{\bf V}_s\over\partial t}+\rho_s({\bf V}_s\cdot
 \nabla){\bf V}_s = - {\rho_s\over\rho}\nabla p_s +
\rho_s S  \nabla T - {\bf F}_{ns}  +\rho_s {\bf T}. %
\ee
 In these equations $p_n$ and $p_s$ are the
effective pressures acting on the normal and the superfluid
component, respectively, defined as $\nabla p_n=\nabla p +( \rho_s/2
) \nabla V_{ns}^2 $ and $\nabla p_s=\nabla p -(\rho_n/2 ) \nabla
V_{ns}^2 $, $p$ the total pressure, $S$ the entropy, $T$ the
absolute temperature, and $\eta$ the dynamic viscosity of the normal
component. The effects of the vortices are described by ${\bf
F}_{ns}$, the friction force exerted by the superfluid component on
the normal component
--- which bears an opposite sign in the evolution equation  (\ref{equa-Vs})
---  and $\rho_s{\bf T}$, the vortex tension force, related to the average
curvature of the vortices, and which is a restoring force arising
in curved vortex lines, as a consequence of the fact that the
streamlines on the side towards the centre of curvature are closer
together than in the external side, which makes a difference of
pressure between both sides, producing and inward-directed force
normal to the filament. Equations (\ref{equa-Vn})-(\ref{equa-Vs})
can be applied when the line length $L$ per unit volume is high,
so that the fluctuations become statistically irrelevant.

The expression  of the mutual friction force ${\bf F}_{ns}$ in the
HVBK equations is \cite{Henderson-TCFD18-2004}
\be\label{Fns-HVBK}%
{\bf F}_{ns}^{(HVBK)}=\rho_s \alpha {\hat\omega}\times[
\vec{\omega}\times ( {\bf V}_{ns}-\tilde\beta \nabla \times  {\bf
\hat{\omega}})] +\rho_s \alpha' \vec{\omega} \times ( {\bf
V}_{ns}-\tilde\beta \nabla \times  {\bf
\hat{\omega}} ), %
\ee
 where $\vec{\omega}=\nabla\times {\bf V}_s$ is the local averaged
superfluid vorticity, $\hat{\omega}=\vec{\omega}/ |\omega|$ the
unit vector along $\vec{\omega}$, and $\tilde\beta$ the vortex
tension parameter, defined as \cite{Don-book}
\be\label{tilde-beta}%
\tilde\beta = {\epsilon_V\over \kappa\rho_s}= {\kappa \over 4\pi}
\ln\left({c\over a_0 L^{1/2}}\right), %
\ee with $c$ a constant of the order of unity, $a_0$ the radius of
the vortex core, of about 1 \AA, and $\epsilon_V$ the energy per
unit length of vortex line. Coefficients $\alpha$ and $\alpha'$
depend on temperature and describe the interaction between the
normal fluid and the vortices. They are linked to the well-known
Hall-Vinen coefficients $B$ and $B'$ by the relations
$\alpha=B({\rho_n/ 2\rho})$, $\alpha'=B'({\rho_n/ 2\rho})$
\cite{Don-book}. Note that this friction force is mediated through
the presence of the vortex lines, as $\vec{\omega}$ is related to
them.

In (\ref{equa-Vs}) $\rho_s{\bf T}$ is the vortex tension force
whose microscopic meaning will be discussed in Section 3. In the
HVBK equations ${\bf T}$ is substituted with
\be\label{T-HALL}%
{\bf T}^{(HVBK)} =(\tilde\beta \nabla\times {\bf\hat\omega})\times
\vec{\omega}
= \tilde\beta (\vec{\omega}\cdot\nabla){\hat\omega}.%
\ee
 From this expression it follows that in this
approximation ${\bf T}$ vanishes when the vorticity is homogeneous
($\nabla \vec{\omega}=0$), or when the vortex lines are parallel
to each other, in which case $\nabla\times {\bf\hat\omega}=0$.

In the  regular vortex array produced by the pure rotation, the
vorticity $\vec{\omega}$ equals $2{\bf \Omega}$ everywhere, ${\bf
\Omega}$ being the angular velocity, and the mutual friction force
${\bf F}_{ns}$ (\ref{Fns-HVBK}) assumes the Hall-Vinen expression
\be\label{Fns-HVBK-Rota}%
{\bf F}_{ns}=2{\rho_s}\alpha {\bf \hat\Omega}\times[{\bf
\Omega}\times
 ({\bf V}_n-{\bf V}_{s})] + 2{\rho_s}\alpha'  {\bf \Omega}\times
 ({\bf V}_n-{\bf V}_{s}),%
 \ee
 because $\nabla\times \vec{\omega}=0$, and ${\vec\omega}=2{\bf \Omega}$.
As seen in (\ref{T-HALL}), the vortex tension $\bf T$ vanishes,
because the vortices are straight lines.

In counterflow superfluid turbulence the vortex tangle is supposed
isotropic  and the mutual friction force (\ref{Fns-HVBK}) is
expressed by \cite{Don-book}--\cite{Vinen-JLTP128-2002}
\be\label{Fns-HVBK-Contro}%
{\bf F}_{ns} = -{2\over 3} \rho_s\kappa\alpha L {\bf V}_{ns}.%
 \ee
The assumed isotropy of the tangle implies that the vortex tension
$\bf T$ vanishes because the microscopic vorticity would be
isotropically distributed, in such a way that the average
${\vec\omega}$ is null everywhere. However, the hypothesis of
complete isotropy of the tangle is not confirmed by the simulation
of the dynamics of the vortex tangle made by Schwarz
\cite{Schwarz-PRL49-1982}--\cite{Schwarz-PRB38-1988} and in the
experiments \cite{Wang-PRB36-1988}. Indeed the particular
direction of the counterflow breaks the rotational symmetry in the
dynamics.  In Section 5, we will consider in detail this case.

The HVBK equations (\ref{equa-Vn})-(\ref{equa-Vs}) have been used
in Refs.
\cite{Swanson-PRL67-1991}--\cite{Hender-PhysB184-2000,Henderson-TCFD18-2004}
to study different experimental situations in liquid He II, as
Taylor-Couette flows, end effects in rotating helium and plane
Poiseuille flow. Recently, these equations have also been used in
Refs \cite{Peralta-AJ635-2005,Peralta-AJ651-2006} to study the
jump in the rotational speed observed in neutron stars, whose
interior is a neutron superfluid at a huge density. In some cases
these equations, with the expressions  of ${\bf F}_{ns}$ and $\bf
T$, were able to describe, in reasonable qualitative agreement
with experiments, the instability of the laminar regime. But,
owing to the limitations imposed by the choice (\ref{Fns-HVBK})
and (\ref{T-HALL}) for ${\bf F}_{ns}$ and $\bf T$, the
applications to other situations is not possible or does not
produce accord with experiments.

One of the aims of the present paper is the generalization of the
expressions (\ref{Fns-HVBK}) and (\ref{T-HALL}) of  ${\bf F}_{ns}$
and $\bf T$, to allow the application of the HVBK equations to
other interesting phenomena. First, we will analyze the limit of
validity of expressions (\ref{Fns-HVBK}) and (\ref{T-HALL}) for
${\bf F}_{ns}$ and ${\bf T}$ in the HVBK model, which, as we will
show in the following Section, are restricted to regions with a
high density of vortex lines, all pointing in the same direction,
and with the same curvature vector. Thus we will pay special
attention to the fluctuations in the direction of the vortex lines
and to the local anisotropy and polarization of the tangle as well
as on its inhomogeneity, and their consequences on ${\bf F}_{ns}$
and ${\bf T}$.

Another topic addressed in the work is to recognize the necessity
to add to equations (\ref{equa-Vn}) and (\ref{equa-Vs}) an
additional equation for vortex line density $L$. In Subsections
4.2 and 4.3 an evolution equation for $L$ is written, which
generalizes Vinen equation to anisotropic, polarized,
inhomogeneous situations.

\section{Expressions for ${\bf F}_{ns}$ and $\bf T$ }
\setcounter{equation}{0}

The HVBK equations (\ref{equa-Vn})--(\ref{equa-Vs}), with ${\bf
F}_{ns}$ and $\bf T$ expressed by (\ref{Fns-HVBK}) and
(\ref{T-HALL}), have been used in several problems in superfluid
hydrodynamics, for instance, to study the Taylor-Couette flow in
helium II
\cite{Swanson-PRL67-1991}--\cite{Hender-PhysB184-2000,Henderson-TCFD18-2004}
and, recently, the jump in the rotational speed observed in
neutron stars \cite{Peralta-AJ635-2005,Peralta-AJ651-2006}. The
numerical simulations obtained with these equations show the
formation of macroscopic vortices, but they do not include the
microscopic chaotic dynamics expressed, for instance, by the
fluctuations of the tangents to the vortex lines.

Our aim here is to incorporate inhomogeneities, anisotropy,
polarization, and tension of the tangle of vortex lines in an
extension of the HVBK  hydrodynamic equations. To this purpose, we
make a critical analysis of these equations, and we perform a
suitable modification of them. Furthermore, we include an
additional equation for the evolution of $L$, besides the
equations (\ref{equa-Vn}) and (\ref{equa-Vs}), because $L$ is
related to forces and tensions.

\subsection{Microscopic expressions for friction force and vortex
tension}

To derive the expressions (\ref{Fns-HVBK}) and (\ref{T-HALL}) of
${\bf F}_{ns}$  and of ${\bf T}$, used in the HVBK equations, and
to determine their limit of validity, we consider the microscopic
description of a vortex tangle in the vortex filament model by
Schwarz \cite{Schwarz-PRL49-1982}--\cite{Schwarz-PRB38-1988}. In
this model, a quantized vortex filament is assumed as a classical
vortex line in the superfluid with  a hollow core and quantized
circulation $\kappa$. The vortex line is described by a vectorial
function ${\bf s}(\xi,t)$, $\xi$ being the arc-length, ${\bf
s'}=\partial{\bf s}/\partial\xi$  the unit vector tangent along
the vortex line, and ${\bf s''}=\partial^2{\bf s}/\partial\xi^2$
the curvature vector. The normal component reacts to a moving
vortex by producing  the "microscopic" mutual friction force
$-{\bf f}_{MF}$. The microscopic form of this force is directly
related to roton scattering from an element of line. The drag
force on unit length of line due to scattering of excitations is
\cite{Don-book}
 \be\label{A1}
 {\bf f}_{D}=- D{\bf s'}\times[{\bf
s'}\times ({\bf v}_R-{\bf v}_{L})] +D_t {\bf s'}\times({\bf
v}_{R}-{\bf v}_{L}) \ee
 where ${\bf v}_R$ is the roton drift velocity near the core and where
the microscopic parameters $D$ and $D_t$ are related to the
kinetic scattering coefficients $\sigma_{\parallel}$ and
$\sigma_{\perp}$ by \cite{Don-book}
 \be\label{A2}D=
\rho_nv_G\sigma_{\parallel} \hskip0.5in D_t=D'-\rho_n \kappa
=\rho_nv_G\sigma_{\perp}-\rho_n \kappa \ee
 where $v_G$ is the thermal average group velocity of rotons.
The last term in equation (\ref{A2}) is due to the presence of the
Iordanskii force \cite{Iordanskii-1965}.

Inserting in (\ref{A1}) the expression of the vortex line element
velocity
 \be\label{A3}{\bf v}_L={d{\bf s}\over dt}={\bf v}_{sl}+\alpha {\bf s'}\times
({\bf v}_n-{\bf v}_{sl}) -\alpha'{\bf s'} \times [{\bf s'} \times
({\bf v}_n- {\bf v}_{sl})],  \ee
 where  ${\bf v}_{sl}=
{\bf v}_{s}+{\bf  v_i}$ is the "local superfluid velocity", sum of
the superfluid velocity at large distance from any vortex line and
of the self-induced velocity --- described by the tangent unit
vector $\bf s'$ and by the curvature $\bf s''$---  the drag force
can be equally described in terms of $({\bf v}_n-{\bf v}_{sl})$ as
follows \cite{Don-book}
\be\label{fms-micro}%
{\bf f}_{MF}= \alpha\rho_s\kappa {\bf s'}\times[{\bf s'}\times
({\bf v}_n-{\bf v}_{sl})] + \alpha'\rho_s\kappa {\bf s'}\times
({\bf v}_n-{\bf v}_{sl}). %
\ee
 Coefficients $\alpha$ and $\alpha'$ are dimensionless
 quantities whose link  to the kinetic scattering coefficients $D$ and $D'$ can be
 found in Ref. \cite{Don-book}.

In the "local induction approximation", that is ignoring
contribution on the self-induced velocity coming from the nonlocal
portion of the vortex, the self-induced velocity ${\bf v_i}$ is
approximated by \cite{Don-book}--\cite{Vinen-JLTP128-2002}
\be\label{vi-loc}%
{\bf v_i}\simeq {\bf v_i}^{(loc)}= \tilde\beta \left[ {\bf s'}\times
{\bf s''} \right].%
\ee The intensity of $\bf v_i$ is $\vert{\bf v_i}\vert \simeq
{\tilde\beta/ R}$, with $R$ the curvature radius of the vortex
lines. The self-induced velocity is zero if the vortices are
straight lines. Observing that ${\bf v}_n-{\bf v}_{sl}={\bf
v}_{ns}-{\bf v_i}$, being ${\bf  v}_{ns}$ the microscopic
counterflow velocity, the mutual friction force {\it per unit
length} assumes the expression
\be\label{fms-micro2}%
{\bf f}_{MF}= \alpha\rho_s\kappa {\bf s'}\times[{\bf s'}\times
({\bf  v}_{ns} -{\bf v_i})] + \alpha'\rho_s\kappa {\bf s'}\times(
{\bf  v}_{ns} -{\bf v_i}).%
\ee

In the evolution equations for the velocities of normal and
superfluid components, the "macroscopic" mutual friction force ${\bf
F}_{ns}$ {\it per unit volume} which superfluid and normal
components mutually exert, is obtained by averaging
(\ref{fms-micro2}) over a small volume $\Lambda$; one has
\be\label{fms-media}%
<{\bf f}_{MF}>= { \int {\bf f}_{MF}
 d\xi \over \int
 d\xi}= {1\over \Lambda L} \int {\bf f}_{MF}
 d\xi, %
\ee where the integral is made over all the vortices contained in
the volume $\Lambda$; to obtain ${\bf F}_{ns}$, we must multiply the
average (\ref{fms-media}), which denotes the average mutual friction
force per unit length, by $ L$, which denotes the length of vortex
lines per unit of volume. One obtains
\be\label{Fns-media}%
{\bf F}_{ns}=[{\bf f}_{ns}]_{av}=L<{\bf f}_{MF}>=
\alpha\rho_s\kappa L<{\bf s'}\times[{\bf s'}\times ({\bf  v}_{ns}
-{\bf v_i})]> + \alpha'\rho_s\kappa L <{\bf s'}\times({\bf v}_{ns}
-{\bf v_i})> .%
\ee

The vortex tension force $\bf T$ arises from the microscopic form
of the evolution equation of ${\bf v}_s$. This equation is a
consequence of the phase-slip of the superfluid wave function
$\varphi$ and, neglecting the mutual friction force, it is written
\cite{Don-book}, \cite{Anderson}
\be\label{vs-dt}%
{\partial {\bf v}_s \over \partial t}= {\bf
v}_L\times \vec{\omega}_{micr} +\nabla \mu.%
\ee
 Here ${\bf v}_L$ is the velocity of the vortex line and the
term $\nabla \mu$ describes the effects of pressure and
temperature gradients,  $\nabla \mu =-\nabla p_s /\rho+S \nabla
T$; ${\bf v}_s$ denotes the microscopic velocity of the superfluid
around the vortex and $\vec{\omega}_{micr}=\oint {\bf v}_s \cdot
d{\bf x}$ the vorticity of the single vortex line.

In equation (\ref{vs-dt}), $\vec{\omega}_{micr}$ is given and it
determines the evolution of ${\bf v}_s$; when making the average
over a volume, $\vec{\omega}_{micr}$  will become a vector field
corresponding to the averages on vortex lines and the
corresponding equation will describe the evolution of the averaged
superfluid velocity.

Note that $\vec{\omega}_{micr}$ is a vector tangent to the vortex
line which describes the quantized vorticity around a vortex line.
On the vortex line itself ${\bf v}_s$ would be singular, but the
rotational is not singular, but is given by the vorticity quantum
$\kappa$.

Note that ${\bf v}_L= {\bf v}_{sl}={\bf v}_s+{\bf v_i}$
\cite{Don-book}, with $\bf v_i$ the "self induced-velocity".
 When the average on the mesoscopic volume $\Lambda$ is taken one
 has
\be\label{vs-dt-macro}%
{\partial {\bf V}_s \over \partial t} = [{\bf
v}_s\times\vec{\omega}_{micr}]_{av}+[{\bf v_i}\times
\vec{\omega}_{micr}]_{av}-{\nabla p_s \over\rho}+S \nabla T, \ee
where $\nabla \mu =-\nabla p_s /\rho+S \nabla T$.

In the HVBK equation, one make the approximation $[{\bf
v}_s\times\vec{\omega}_{micr}]_{av}= {\bf V}_s\times({\nabla\times
{\bf V}_s})= -({\bf V}_s\cdot\nabla){\bf V}_s +(1/2) \nabla {\bf
V}_s^2$, obtaining
\be\label{ro-vs-dt}%
\rho_s{\partial {\bf V}_s \over \partial t}+\rho_s({\bf
V}_s\cdot\nabla){\bf V}_s =\frac{1}{2}\nabla {\bf V}_s^2+ \rho_s
{\bf T} - {\rho_s\over\rho}\nabla p_s+\rho_sS \nabla T .%
\ee
 This allows to identify the tension $\bf T$ as
\be\label{T-macro}%
{\bf T}= [{\bf v_i} \times \vec{\omega}_{micr}]_{av}= \kappa
L<{\bf v_i} \times{\bf s'}> ,%
\ee
 where we have denoted the average of
${\bf v_i} \times \vec{\omega}_{micr}$, over the small vortex
tangle contained in the mesoscopic volume $\Lambda$, with angular
brackets, as in (\ref{fms-media}). Equation (\ref{T-macro}) makes
explicit that the tension force $\bf T$ is related to the average
curvature of vortex lines, as it was mentioned above.

 Relation (\ref{T-macro}) may be rewritten in several
equivalent forms, by taking into account (\ref{vi-loc}) and some
vectorial identities. For instance, $<{\bf s'}\times{\bf v_i} >$
may be expressed, in the local induction approximation,  as
\be\label{sPrim-vi}%
<{\bf s'}\times{\bf v_i} >= \tilde\beta<{\bf
s'}\times ({\bf s'}\times{\bf s''} )
> =- \tilde\beta<{\bf s''}>,%
\ee where it has been taken into account that ${\bf s'}$ is a unit
vector.

Using (\ref{sPrim-vi}), expression (\ref{T-macro}) may be
rewritten as
\be\label{T-macro-av}%
{\bf T}=[{\bf v_i} \times \vec{\omega}_{micr}]_{av}=\kappa L
\tilde\beta <({\bf s'\times s''}) \times {\bf s'}>= \kappa L
\tilde\beta <{\bf s''}>.%
\ee
 Relation (\ref{T-macro-av}) is especially interesting because it
explicitly describes the connection between $T$ and the curvature
of the vortex lines, which is given by $\bf s''$.

 Another useful form for ${\bf T}$ may be obtained by using
\be\label{sSec}%
{\bf s''}= ({\bf s'\cdot \nabla}){\bf s'}= -{\bf s'}\times
(\nabla\times {\bf s'})%
\ee so that expression (\ref{sPrim-vi}) may be rewritten as
\be\label{sPrim-vi2}%
<{\bf s'}\times{\bf v_i} >=\tilde\beta<{\bf s'} \times \left({\bf
s'}\times {\bf s''}\right)> = \tilde\beta <{\bf s'}\times
(\nabla\times {\bf s'})>. %
\ee
 Expressions (\ref{T-macro}), (\ref{T-macro-av}) and (\ref{sPrim-vi2}) will be used in the next sections.

\subsection{Limit of validity of the  HVBK's equations}
 As it can be seen by comparing (\ref{Fns-HVBK}) and (\ref{Fns-media}), in expression (\ref{Fns-HVBK}) of the mutual
friction force ${\bf F}_{ns}$  used in the HVBK equations the
quantities $<{\bf s'}\times({\bf s'}\times ({\bf v}_{ns}-{\bf
v_i}) )>$ and $<{\bf s'}\times({\bf v}_{ns} -{\bf v_i})>$ are
approximated with ${\hat\omega}\times[ \hat{\omega}\times ( {\bf
V}_{ns}-\tilde\beta \nabla \times  {\bf \hat{\omega}})]$ and ${\bf
\hat\omega}\times ( {\bf V}_{ns}-\tilde\beta \nabla \times  {\bf
\hat{\omega}} )$ respectively,  and $|\nabla \times {\bf v}_s|$ is
approximated by $\kappa L$. The same approximations, when applied
to the vortex tension $\bf T$, yield $(\tilde\beta \nabla \times
{\bf \hat{\omega}}) \times \vec{\omega}$.

We analyze in this section the limits of validity of the
approximations leading to these expressions for 
${\bf F}_{ns}$  and 
 ${\bf T}$, and we will show that
these limits are related to the neglect of the second order
moments of the vector $\bf s'$. We consider first the two averaged
quantities $<{\bf s'}\times({\bf s'}\times {\bf v}_{ns} )>$ and
$<{\bf s'}\times {\bf v}_{ns}>$.
 We denote with ${\bf p}=<\bf  s'>$ and ${\bf V}_{ns}$
the averaged values  of ${\bf s'}$  and ${\bf v}_{ns}$, and with
$\delta{\bf s'}$ and $\delta{\bf v}_{ns}$ their respective
fluctuations. The average value of the unit vector $\bf s'$ is
called {\it tangle polarity} \cite{Jou-PRB74-2006}--
\cite{Barenghi-PRL89-2002}, and
--- recall that the superfluid vorticity is quantized --- is
linked to the local averaged superfluid vorticity $\vec \omega$ by
the relation
\be\label{p-def}%
{\bf p}= <{\bf  s'}>=
{1\over \Lambda L} \int {\bf s'}
 d\xi = {{\vec \omega}\over\kappa L}= {\nabla \times
{\bf V}_s\over\kappa L}.%
\ee
 For a totally polarized tangle with all the tangents $\bf s'$
parallel to each other, $|{\bf p}|=1$. In two dimensions, the
polarization may be interpreted as $(n^+-n^-)/(n^++n^-)$ with
$n^+$ the vortices rotating in one direction and  $n^-$ the
vortices rotating in opposite direction. Processes producing a
partial separation of $+$ and $-$ vortices will thus change $\bf p$.
This means that vortices may be partially agglomerated in such a
way that $+$ vortices will be predominant in some zones to the
detriment of other ones, where $-$ vortices will be. This leads to
the formation of relatively large $+$ and $-$ vortices by local
macroscopic accumulation of the corresponding $+$ and $-$ microscopic
quantized vortices \cite{Zhang-vanSciver}.  Thus, the dynamics of
$\bf p$ may be certainly important and it can be influenced by the
boundary conditions.

With the notation just introduced, neglecting the fluctuations of
the counterflow velocity ${\bf V}_{ns}$, we may write ${\bf s'}={\bf
p} +\delta {\bf s'}$. Since $<\delta{\bf s'}>=0$, we obtain
\be\label{sPrim-fluttu-1}%
 <{\bf s'}\times({\bf s'}\times{\bf v}_{ns})> ={\bf
p}\times ({\bf p}\times {\bf V}_{ns})+<{\bf \delta s'}\times ({\bf
\delta s'}\times {\bf  V}_{ns})>,%
\ee
\be\label{sPrim-fluttu-2}%
<{\bf s'}\times {\bf v}_{ns} >= {\bf p}\times{\bf V}_{ns}.%
\ee  Note that in the equation (\ref{Fns-HVBK}) only the first
terms in the right-hand side of equations (\ref{sPrim-fluttu-1})
and (\ref{sPrim-fluttu-2}) appear. Indeed, recalling equation
(\ref{p-def}), the first terms in the right-hand side of equations
(\ref{sPrim-fluttu-1}) and (\ref{sPrim-fluttu-2}) can be written
as
\be\label{p-vetto}%
 {\bf p}\times ({\bf p}\times {\bf
V}_{ns})={1\over \kappa L} {\hat\omega}\times [ \vec{\omega}
\times   {\bf V}_{ns}]= <{\bf s'}\times({\bf s'}\times{\bf
v}_{ns})>^{(HVBK)},%
\ee
\be\label{p-vetto-2}%
{\bf p}\times {\bf V}_{ns} ={1\over \kappa L} \vec{\omega}\times
{\bf V}_{ns}= <{\bf s'}\times{\bf v}_{ns}
>^{(HVBK)}, \ee
 whereas the other terms, quadratic in the fluctuations, have been
 neglected. In general, this will not be correct, for instance, in
 the limiting situations of an isotropic tangle, ${\bf p}=0$, but
$<{\bf s'}\times({\bf s'}\times{\bf v}_{ns})>= (2/3){\bf V}_{ns}$
\cite{Jou-PRB74-2006}.

Note also that in this simplified hypothesis, the second term in
equation (\ref{sPrim-fluttu-1}), dependent on the second moments
of $\bf s'$,  can be neglected only if the orientational
fluctuations of this unit vector in the small volume $\Lambda$ are
very small, i.e. if most of the vortex lines in the volume have
the same direction.

Furthermore,  making use of equation (\ref{sSec}), we obtain
\be\label{vi-loc-medio}%
<{\bf v_i}> =\tilde\beta< {\bf
s'}\times{\bf s''}> = \tilde\beta<(\nabla\times {\bf s'})- [{\bf
s'}\cdot (\nabla\times {\bf s'})]{\bf s'}> =\tilde\beta<({\bf
U}-{\bf s's'})\cdot (\nabla\times {\bf s'}) >,
 \ee
 with $\bf U$ the unit matrix and $\bf s's'$ the diadic product.
Note that in the HVBK equations the quantity $<{\bf v_i}>$ is
simply approximated by $\tilde\beta <\nabla \times {\bf s'}>$ and
therefore the quantity $<[{\bf s'}\cdot (\nabla\times {\bf
s'})]{\bf s'}> $ is neglected.

From equations (\ref{p-vetto})--(\ref{p-vetto-2}) it is seen that
in the HVBK equations all the second-order moments of the
fluctuations in $\bf s'$ are neglected, that is $\bf s'\simeq
<{\bf s'}>={\bf p}$, which implies $|<{\bf s'}>|\simeq 1$ or
\be\label{kappa-L}%
\kappa L \simeq |\nabla \times {\bf V}_s|= |\vec{\omega}|.%
\ee
 Last relation is the most critical hypothesis when it is used
to evaluate the line density $L$. In fact, if in a mesoscopic
region $\Lambda$ there are several vortex lines oriented in a
random way, the line density $L$ in $\Lambda$ will be very
different from $L \simeq {|\nabla \times {\bf V}_s|/ \kappa }$
--- which corresponds to an extreme polarization with all or
almost all the lines pointing out in the same direction. We stress
on this statement by a trivial example: let us consider in a small
region $\Lambda$ a single vortex loop, then the average $|<{\bf
s'}>|$ on this loop is zero, but this is not the case for the
vortex line density $L$.

As a consequence, the HVBK equations  describe correctly the
interaction between the normal component and the vortex tangle
only in mesoscopic regions with a high density array of vortex
lines, all pointing in the same direction (totally polarized) and
with the same curvature vector. This fact  would   explain the
results of the numerical simulations obtained using the HVBK
equations, which show the formation of macroscopic vortices and in
which the microscopic chaotic dynamic behavior of the vortices
does not appear.  Of course, a confirmation of our statement
should come from the results of numerical simulations of our
equations and of the HVBK equations to problems with vortices not
completely polarized.

\section{Generalization of HVBK equations}
\setcounter{equation}{0} In a previous paper
\cite{Mongiovi-PRB75-2007} a hydrodynamical model of turbulent
superfluids was formulated, which uses as fundamental fields the
mass density ${\rho}$, the velocity ${\bf v}$ of the helium as a
whole, the temperature $T$, the heat flux ${\bf q}$ and the line
density $L$. In that work, only situations in which the tangle can
be supposed approximately isotropic were considered, as is often
made in the study of counterflow superfluid turbulence. In Ref.
\cite{Jou-PRB74-2006} the anisotropy of the vortex tangle was
studied, restricting the study to homogeneous situations, and
neglecting the influence of the vortex tension. In another paper
\cite{Sciacca-Couette} we have considered a more general
situation, taking into account of inhomogeneities, anisotropy and
polarization of the vortex tangle, studying the plane Couette and
Poiseuille flow. For the sake of simplicity, in that work, we have
neglected the vortex tension $\bf T$.

\subsection{New  determination of ${\bf F}_{ns}$ and $\bf T$}

We want now to obtain expressions for ${\bf F}_{ns}$ and $\bf T$
overcoming some of the restrictions mentioned in the Section 3.2.
Using the local-induction approximation, that is ignoring
contribution on the self-induced velocity coming from the nonlocal
portion of the vortex, and neglecting the fluctuations of the
relative velocity ${\bf V}_{ns}$, from (\ref{Fns-media}) we deduce
that ${\bf F}_{ns}$ can be written \cite{Jou-PRB74-2006}
\be\label{Fns-new}%
{\bf F}_{ns}=-\rho_s\kappa L\left[ \alpha <{\bf U} - {\bf s'\bf
s'}>  + \alpha'  <{\bf W} \cdot {\bf s'}>\right]\cdot   {\bf
V}_{ns} +\rho_s\kappa L\tilde\beta\left[ \alpha <{\bf s'\times
s''}>+\alpha' <{\bf s''}>\right], \ee
 with ${\bf W}$ the Ricci tensor (a completely antisymmetric third-order
tensor such that ${\bf W} \cdot {\bf s'}\cdot  {\bf V}_{ns}=-{\bf
s'}\times {\bf V}_{ns}$).

Introducing the tensor ${\bf\Pi}={\bf\Pi}^s+{\bf\Pi}^a$,
 with \cite{Jou-PRB74-2006}
\be\label{Pi-s}%
{\bf\Pi}^s  \equiv{3\over 2} <{\bf U} -{\bf s'\bf s'}>,
\hskip0.3in {\bf\Pi}^a \equiv{3\over 2}{\alpha'\over\alpha} <{\bf
W}\cdot {\bf s'}> , \ee the vectors $\bf I$ and $\bf J$
\cite{Schwarz-PRB38-1988}, \cite{Jou-PRB74-2006}
\be\label{I-and-J}%
{\bf I} \equiv{ \int {\bf
s'}\times {\bf s''} d\xi \over \int | {\bf s''} | d\xi },
\hskip0.4in
 {\bf J} \equiv { \int {\bf s''} d\xi
\over \int | {\bf s''} | d\xi}, \ee
 and  $c_1L^{1/2}={1\over \Lambda L} \int | {\bf s''} | d\xi$, a
characteristic measure of the vortex tangle introduced by Schwarz
\cite{Schwarz-PRB38-1988}, the mutual friction force (\ref{Fns-new})
can be written in a compact way as
\be\label{Fns-new-2}%
{\bf F}_{ns}=  \alpha\rho_s\kappa L \left[-{2\over 3}
 {\bf \Pi} \cdot  {\bf V}_{ns} +
\tilde\beta c_1L^{1/2} \left(  {\bf {   I}}  + {\alpha'\over
\alpha} {\bf J} \right) \right] . \ee
 The tensor $\bf \Pi$ is especially useful to describe
the geometrical properties of the tangle related with the
orientational distribution of the vortex lines, whose local
direction is indicated by the unit tangent $\bf s'$.  When the
tangle is not completely polarized nor fully isotropic, it should
be described by a tensor, rather than by a vector, as it is
usually done in the description of liquid crystals and of polymer
solutions, where a tensor similar to ${\bf \Pi}^s$ in (\ref{Pi-s})
is used \cite{Doi},\cite{Ottinger}. The tensor $\bf\Pi$ does not
contain information on the curvature of the lines because it does
not contain $\bf s''$; this is furnished by the vectors ${\bf I}$
and ${\bf J}$.

 The vortex tension $\bf T$ is also linked to the curvature vector $\bf
 J$; indeed it results
\be\label{T-new}%
{\bf T}= [{\bf v_i}\times (\nabla\times {\bf v}_s)]_{av} =\kappa L
<(\tilde\beta {\bf s'}\times {\bf s''})\times {\bf s'}> = \kappa
L\tilde\beta  <{\bf s''}> = \kappa L^{3/2} c_1{\bf J}. \ee

Observe that the tension of the vortex line, as the curvature
vector, is zero in pure rotation with parallel straight vortex
lines and in the isotropic tangle produced in well developed
counterflow turbulence, but it is not so in the presence of
simultaneous counterflow and rotation
\cite{Tsubota-PRB69-2004,Jou-PRB69-2004} or in the first stages of
the turbulence
\cite{Mongiovi-PRB75-2007-214514,Vinen-PRSLA240-1957} or in the
transient states after sudden acceleration in plane Couette and
Poiseuille flow. For instance, ${\bf T}$ could be different than
zero in Kelvin helical vortex waves, which appear when a
sufficiently intense counterflow is superposed to an axial
rotation. In the transition from the array of straight lines to
the array of helical vortex lines, a spiral tension would appear.
The tension ${\bf T}$ would be a rotating quantity, as well as the
curvature ${\bf s''}$ of the helical vortices. If the wavelength
becomes shorter than the characteristic observational length, the
average ${\bf T}$ will become equal to zero, but if the wavelength
is high enough, it could lead to specific secondary flows of ${\bf
v}_s$, through equation (\ref{equa-Vs}).

As a first modification to the HVBK equations we propose to take
into account the fluctuations of the vector $\bf s'$, which appear
in the tensor $\bf <U- s's'>$, but, for sake of simplicity, we
propose to neglect the fluctuations of $ 
\bf s''$. This can be made following several way. Note indeed that
in expressions (4.1) and (4.5) the quantity $<\bf s''>$ appears,
which is in general independent from $<\bf s'>= p$ and from $<\bf
s's'>=U- \Pi $. However, our aim is to formulate an hydrodynamical
model for a superfluid, with the lowest possible number of unknown
quantities. As a consequence, we can try to find an approximate
expression of $<\bf s''>$ in terms of  $\bf p$, $\bf \nabla p$ and
$\bf \Pi$; the presence of $\bf \nabla  p$ is logical because $\bf
s''$ is related to the spatial derivative of $\bf s'$; thus,
although our expression will be only an approximate one, it makes
clear that $\bf p$ and $\bf \Pi$ by themselves are not sufficient
to give a description of $\bf s''$.  This may be made, taking in
mind relations (3.15). Indeed, putting $\bf <s'>= p + \delta s'$,
we have:
\be\label{A.1}%
<{\bf s''}>=-<{\bf s'\times(\nabla\times s')}> = -{\bf
p\times(\nabla\times p) - <\bf \delta s'\times(\nabla\times \delta
s')>}  \ee
 or also, noting that $\bf \nabla \cdot (s's')= (s' \cdot \nabla)s'+
 (\nabla\cdot s') s'$:
\be\label{A.2}%
<{\bf s''}>=<{\bf (s' \cdot \nabla)s'}> =<{\bf \nabla \cdot
(s's')- (\nabla\cdot s')s'}>= - {\nabla {\bf \Pi}}-{\bf (\nabla\cdot
p)p} - <{\bf(\nabla\cdot \delta s')\delta s'}>  \ee
 As one sees, using the approximation
\be\label{A.3}%
<{\bf s''}> \simeq -{\bf p\times(\nabla\times p)}\ee
 one neglect the quantity ${<\bf \delta
s'\times(\nabla\times \delta s')>}$, i.e. an average of vectors
orthogonal to $\delta \bf s'$, while, if one puts
\be\label{A.4}%
<{\bf s''}>  \simeq - {\bf\nabla \Pi}-{\bf (\nabla\cdot p)p}\ee
 one neglect the quantity
$<{\bf(\nabla\cdot \delta s')\delta s'}>$, i.e. an average of
vectors collinear to $\delta \bf s'$.

The first approximation is true for very polarized or for
completely non-polarized situations; in the first one, almost all
tangent vectors to the lines go in the same direction, whereas in
the last situation all of them go in different directions, in such
a way that they cancel each other and yield null polarization. The
second approximation is more useful in intermediate situations.
Another possibility, which may be useful in situations with a high
but not total polarization,  would be to use an average of the two
previous approximated expressions (\ref{A.3}) and (\ref{A.4}), but
we will not deal with it for the sake of simplicity. In the
following we will use for $<\bf s''>$ the approximation
(\ref{A.3}).

To what concerns the vectors ${\bf I}$ and ${\bf J}$, and the
vortex tension $\bf T$, with these approximations in mind, we have
 the following constitutive relations
\be\label{c1LI}%
c_1 L^{1/2}{\bf I} =<{\bf s'}\times {\bf s''} >
\simeq <{\bf U} -{\bf s'\bf s'}> \cdot \nabla \times  {\bf p}  ,
\ee
\be\label{c1LJ}%
c_1 L^{1/2}{\bf J} =<{\bf s''}> \simeq -{\bf p} \times (\nabla \times
 {\bf p})=({\bf p}\cdot \nabla){\bf p}-{1\over2} \nabla {\bf p}^2 .
 \ee
The required equations for ${\bf F}_{ns}$ and $\bf T$ are
\[
{\bf F}_{ns}=-\rho_s\kappa L   \alpha
<{\bf U} - {\bf s'\bf s'}> \cdot[{\bf V}_{ns}-\tilde\beta
(\nabla\times {\bf p})] + \rho_s\kappa L\alpha' [{\bf p}\times
{\bf V}_{ns} -\tilde\beta {\bf p} \times (\nabla \times
 {\bf p})]=
 \]
\be\label{Fns-new-finale}%
=-\frac{2}{3}\rho_s\kappa L \alpha {\bf \Pi} \cdot[{\bf
V}_{ns}-\tilde\beta (\nabla\times {\bf p})], \ee
\be\label{roT-new-finale}%
\rho_s{\bf T}= - \rho_s\kappa L \tilde\beta {\bf p} \times (\nabla
\times {\bf p}). \ee

In Section 5, some illustration of (\ref{Fns-new-finale}) and
(\ref{roT-new-finale})  will be presented, with different form of
the field ${\bf \Pi}$. But, to complete our hydrodynamical model,
we must first add to equations
(\ref{Fns-new-finale})--(\ref{roT-new-finale}) an evolution
equation for the line density $L$.

\subsection{Generalized Vinen equation including  polarization,
anisotropy, and inhomogeneities} In expressions (\ref{Fns-new})
and (\ref{T-new}) --- or (\ref{Fns-new-finale}) and
(\ref{roT-new-finale}) --- for ${\bf F}_{ns}$ and $\bf T$, it
appears $L$; therefore, to have a full description for the
evolution of the system, an evolution equation for $L$ is needed.

The evolution equation for $L$ in counterflow superfluid
turbulence was formulated by Vinen \cite{Vinen-PRSLA240-1957}.
Assuming homogeneous turbulence, such an equation is
\be\label{deL-su-dt}{dL\over dt}  = \alpha_v V_{ns} L^{3/2}   -
\beta_v \kappa L^2, \ee
 with $\alpha_v$ and $\beta_v$ dimensionless parameters.

A microscopic derivation of this equation was given  by Schwarz
\cite{Schwarz-PRL49-1982,Schwarz-PRB38-1988} on the basis of the
dynamics of the vortices, neglecting however a term in the average
curvature vector ${\bf s''}$. In Ref. \cite{Jou-PLA368-2007},
because we were interested to study wall effects on the evolution
of $L$, the following extension of equation (\ref{deL-su-dt}) was
written \be\label{deL-su-dt2}%
{dL\over dt} \simeq \alpha c_1 {\bf V}_{ns}\cdot{\bf I} L^{3/2} +
{\alpha' }c_1{\bf V}_{ns}\cdot {\bf J}  L^{3/2}- \alpha\tilde\beta
c_2 L^2 , \ee
 with $c_1$, $\bf I$ and $\bf J$ defined in
equations (\ref{I-and-J}), $c_1$ in the line below
(\ref{I-and-J}), and $c_2= {1\over \Lambda L^{2}} \int | {\bf s''}
|^2 d\xi$.

Substituting in this equation the relations
(\ref{c1LI}--\ref{c1LJ}), the following evolution equation for $L$
which takes into account the polarization and the anisotropy of
the tangle is obtained \be\label{deL-su-dt21}%
 {dL\over dt}
\simeq \alpha L {\bf V}_{ns}\cdot<{\bf U- s's'}> \cdot (\nabla
\times {\bf p})- {\alpha' }L{\bf V}_{ns}\cdot {\bf p}\times
(\nabla \times {\bf p})- \alpha\tilde\beta c_2({\bf p}) L^2 . \ee
In Refs. \cite{Sciacca-Couette},
\cite{Jou-PRB69-2004}--\cite{Mongiovi-PRB75-2007-214514},
\cite{Geust-PhysB154-1989}--\cite{Jou-PRB72-2005-144517} Vinen's
equation was modified to describe more complex situations, as for
instance the coupled situation of counterflow and rotation and in
Couette and Poiseuille flows. Taking in mind the equation
(\ref{deL-su-dt21}) and the proposed equation of Ref.
\cite{Sciacca-Couette}, the previous equation would become
\be\label{deL-su-dt3}%
 {dL\over dt}
\simeq \alpha L {\bf V}_{ns}\cdot<{\bf U- s's'}> \cdot (\nabla
\times {\bf p})- {\alpha' }L{\bf V}_{ns}\cdot {\bf p}\times (\nabla
\times {\bf p})- \alpha\tilde\beta \kappa L^2 \left[1-\sqrt{|{\bf
p}|}\right]\left[1-B\sqrt{|{\bf p}|}\right], \ee where $B$ is a
dimensionless coefficient lower than 1. A rigorous derivation of its
form would require a microscopic extension of Schwarz's model
including rotation effects. To check the consistency of the obtained
evolution equation (\ref{deL-su-dt3}) for the vortex line density
equation, in the next subsection an analysis based on the formalism
of linear irreversible thermodynamics will be made.

When inhomogeneities in the line density $L$ are taken into
account, the evolution equation for line density $L$ must include
a vortex density flux ${\bf J}^L$, as \cite{Mongiovi-PRB75-2007}
\be\label{deL-su-dt-Nabla}%
 {\partial L\over \partial t}  +
\nabla\cdot {\bf J}^L  =\sigma^L, \ee
 where $\sigma_L$ stands for the right-hand side of equation
(\ref{deL-su-dt3}). The flux ${\bf J}^L$ can also be expressed in
terms of a convective part $L{\bf v}^L$  with ${\bf v}^L$  the
tangle velocity and a dissipative part ${\bf J}^L_d$. The particular
form of ${\bf J}^L$ is also open to the debate. One can suppose that
${\bf J}^L$ is an independent variable \cite{Sciacca-MCM-inpress},
or, more simply, supposing that ${\bf J}^L$ is a dependent field: we
have found \cite{Mongiovi-PRB75-2007} for it the form ${\bf
J}^L=\nu_0{\bf q}$, $\nu_0$ being a coefficient describing the
interaction between the vortex tangle and the heat flux $\bf q$,
which is linked to the counterflow velocity by the relation ${\bf
q}=\rho_s TS {\bf V}_{ns}$. The heat flux $\bf q$ may also be
expressed in terms of $\nabla T$
and $\nabla L$ as \cite{Mongiovi-PRB75-2007} \be\label{q-def}%
{\bf q}=- {\eta\over \kappa L_0}\nabla T- {\chi_0\over \kappa L_0
}\nabla L . \ee Thus, the dissipative part of the vortex flux
${\bf J}^L_d$ may be expressed as
\be\label{Jd}%
{\bf J}^L_d=- {\nu_0\eta_0\over \kappa L_0 }\nabla T -
 {\nu_0\chi_0\over \kappa L_0 }\nabla L .
\ee

 In isothermal situations, ${\bf J}^L_d$ may be written as ${\bf J}^L_d=-D\nabla
L$, with  $D$ a vortex diffusion coefficient defined by
$D={\nu_0\chi_0/ \kappa L_0 }$ \cite{Jou-PLA368-2007}.

\subsection{Consequences of Onsager-Casimir reciprocity relation}
In this Subsection, we show that another modification of the vortex
line density evolution equation is necessary to insure the
thermodynamic consistency of the evolution equations for $L$ and for
${\bf V}_s$, according to the formalism of linear irreversible
thermodynamics \cite{Jou-PRB72-2005-144517, de Groot-Amsterdam1962,
G. Lebon-Berlin2008}. This analysis requires the presence of another
term linked to the tension of vortices.

We follow the general lines of \cite{Jou-PRB72-2005-144517, de
Groot-Amsterdam1962, G. Lebon-Berlin2008} with the aim to study
the consequences of the Onsager-Casimir reciprocity relations on
the evolution equations of ${\bf V}_s$ and $L$ proposed in the
previous Sections. The Onsager relations were derived at the
microscopic scale by using fluctuation theory and microscopic time
reversibility, and they do not depend on the details of the
microscopic models; this is the reason that their validity is not
restricted to dilute gases, but it has been checked for different
kinds of systems (dilute and dense gases, liquids, solids,
multicomponent fluids, artificial and biological membranes).

According to the formalism of nonequilibrium thermodynamics one
may obtain evolution equations for ${\bf V}_s$ and $L$ by writing
${d {\bf V}_s/ dt}$ and ${dL/ dt}$ in terms  of their conjugate
thermodynamic forces $-\rho_s {\bf V}_{ns}$ and $\epsilon_V$. The
evolution equation for ${\bf V}_s$, neglecting inhomogeneous
contributions of pressure, temperature and velocity, in an
inertial frame, is written
\be\label{devsdt}%
\rho_s{d {\bf V}_s \over d t}  =-{\bf F}_{ns} - \rho_s {\bf T}=
   \alpha\rho_s\kappa L {2\over3}{\bf\Pi}\cdot {\bf V}_{ns}-
   \alpha\rho_s\kappa L\tilde\beta
{2\over3}{\bf\Pi}\cdot (\nabla \times {\bf p}) + \rho_s\kappa L
\tilde\beta {\bf p} \times (\nabla \times {\bf p})%
\ee
 and the evolution equation for $L$ is written
 \be\label{deL-su-dt2-2}%
 {dL\over dt}
= \alpha L {\bf V}_{ns}\cdot<{\bf U- s's'}> \cdot (\nabla \times
{\bf p})- {\alpha' }L{\bf V}_{ns}\cdot {\bf p}\times (\nabla \times
{\bf p})- \alpha\tilde\beta c_2({\bf p}) L^2 . \ee

However, in the right-hand side of (\ref{deL-su-dt2-2}) additional
contributions  must be included to make (\ref{deL-su-dt2-2})
thermodynamically consistent with (\ref{devsdt}).

Similarly to that presented in \cite{Jou-PRB72-2005-144517}, we
write $d {\bf V}_s /d t$ and $d L/d t$ in matrix form using the
equations (\ref{devsdt}) and (\ref{deL-su-dt2-2}), and  by means of
Onsager-Casimir reciprocity we obtain an additional contribution to
the evolution equation for $L$. The result is
\be\label{matrix}%
\begin{pmatrix}
  {d {\bf V}_s\over dt}   \\
  {d L\over dt}
\end{pmatrix}
= L
\begin{pmatrix}
  - {2\over3}  \frac{\alpha\kappa}{\rho_s} {\bf
\Pi} & -{2\over3} {\alpha\over\rho_s}(\nabla\times {\bf p})
\cdot{\bf \Pi}+ {1\over\rho_s}{\bf p}\times (\nabla
\times {\bf p}) \\
-{\alpha\over\rho_s}{2\over3}(\nabla\times {\bf p}) \cdot{\bf \Pi}+
{1\over\rho_s}{\bf p}\times (\nabla \times {\bf p}) & -
{\alpha\tilde\beta \over
 \epsilon_V}L
 c_2({\bf p})
\end{pmatrix}
\begin{pmatrix}
-\rho_s{\bf V} \\ \epsilon_V
\end{pmatrix}%
\ee
 where for $c_2({\bf p})$ we can choose the expression
 $c_2({\bf p}) =\left(1 -\sqrt{|{\bf p}|}\right) \left(1 -B\sqrt{|{\bf p}|}\right) $
 found in Ref. \cite{Sciacca-Couette}.

Therefore the equation for $d L/dt$ becomes
\be\label{deL-su-dt-Onsager}%
 {dL\over dt}
= \alpha L {\bf V}_{ns}\cdot<{\bf U- s's'}> \cdot (\nabla \times
{\bf p})-(1+ {\alpha' })L{\bf V}_{ns}\cdot {\bf p}\times (\nabla
\times {\bf p})- \alpha\tilde\beta c_2({\bf p}) L^2, \ee or,
similarly,
\be\label{deL-su-dt-Onsager2}%
 {dL\over dt}
= \frac{2}{3}\alpha L {\bf V}_{ns}\cdot {\bf \Pi}\cdot (\nabla
\times {\bf p})-\frac{1}{\kappa \tilde \beta}{\bf V}_{ns}\cdot {\bf
T}- \alpha\tilde\beta \left(1 -\sqrt{|{\bf p}|}\right) \left(1
-B\sqrt{|{\bf p}|}\right) L^2. \ee

The new term not contained in  the evolution equation
(\ref{deL-su-dt2-2}) for $L$ is the coupling term between $d L/dt$
and $ -\rho_s{\bf V}_{ns} $ in the matrix in (\ref{matrix}), which
is linked to the tension of vortices and it is null when the tension
${\bf T}$ is null.

Note that, introducing the tensor
\be\label{new-tensor-Onsager}%
  {\bf \Pi'}=<{\bf U- s's'}> +{1+\alpha'\over\alpha}  <{\bf W} \cdot {\bf s'}>  \ee
the system (\ref{matrix}) can be written
\be\label{matrix-Onsager}%
\begin{pmatrix}
  {d {\bf v}_s\over dt}   \\
  {d L\over dt}
\end{pmatrix}
= L
\begin{pmatrix}
  - {2\over3}  \frac{\alpha\kappa}{\rho_s} {\bf
\Pi} & -{2\over3}{\alpha\over\rho_s}{\bf
\Pi'} \cdot (\nabla\times {\bf p})\\
-{2\over3}{\alpha\over\rho_s}{\bf \Pi'} \cdot (\nabla\times {\bf
p})& - {\alpha\tilde\beta \over
 \epsilon_V}L
 c_2({\bf p})
\end{pmatrix}
\begin{pmatrix}
-\rho_s{\bf V} \\ \epsilon_V
\end{pmatrix}%
\ee

\section{Application to rotating counterflow turbulence}
\setcounter{equation}{0}
 Combination of counterflow and rotation is especially interesting
in the context of the present paper, because it provides
intermediate situations between an isotropic tangle and a totally
anisotropic array of parallel vortices. The polarization and
anisotropy of the tangle depend on the ordering influence of the
rotation, which tend to align the vortices parallel to the rotation
axis, and the randomizing aspects of the counterflow. In this
section we will apply the previous results.

\subsection{Pure counterflow}
 Experimental observations \cite{Wang-PRB36-1988} and numerical simulations
\cite{Schwarz-PRB38-1988} in counterflow superfluid turbulence
show anisotropy in the vortex line distribution with vortices
concentrated in planes orthogonal to ${\bf V}_{ns}$. Assuming
${\bf V}_{ns}$ in the direction of the $x$ axis and isotropy in
planes orthogonal to it, one can choose for the tensor $\bf \Pi$
(equation (\ref{Pi-s})) the following expression
\cite{Jou-PRB74-2006} \be\label{Pi-H}%
{\bf\Pi}_H={\bf\Pi}_H^s={3\over2} \left(
                                      \begin{array}{ccc}
                                        2a& 0& 0 \\
                                        0&1-a&0 \\
                                        0&0&1-a \\
                                      \end{array}
                                    \right); \ee
here $a$ is the anisotropic parameter linked to the coefficients
$I_\|$ and $I_\bot$ introduced by Schwarz
\cite{Schwarz-PRB38-1988} by the relations $I_\|= 2a$ and $I_\bot=
1-a$. Being known the values of coefficient $a$, by accurate
measurements of second sound, the second order moments of the unit
vector $\bf s'$ remain determined. Indeed, it results
\be%
<{s'_x}^2>=1-2a, \hskip0.4in <{s'_y}^2> = <{s'_z}^2>=a .%
\ee

For the vector $\bf I$, we propose, according to (\ref{c1LI}), the
following constitutive relation \be\label{c1LI-2}%
c_1 L^{1/2}{\bf I} =<{\bf
s'\times s''}>\simeq <{\bf U} - {\bf s'\bf s'}> \cdot \nabla
\times
 {\bf {p}},%
 \ee
The asymmetric part of the tensor ${\bf\Pi}$, the curvature vector
$\bf J$ and the vortex tension $\bf T$ in this case are zero,
owing to the supposed isotropy of the tangle in planes orthogonal
to ${\bf V}_{ns}$, which implies $<{\bf s''}>=0$. Introducing
(\ref{Pi-H}) into (\ref{Fns-new-finale}) and (\ref{deL-su-dt3}) we
could obtain the influence of the anisotropy $a$ on the mutual
friction force ${\bf F}_{ ns}$ and the evolution of $L$.

\subsection{Simultaneous counterflow and rotation}.
Under the simultaneous influence of counterflow velocity ${\bf
V}_{ns}$ and rotation with angular speed $\bf \Omega$, rotation
tends to align vortex lines parallel to rotation axis, whereas
counterflow velocity tends to produce a disordered tangle. In
these situations one has partially polarized tangles, requiring
the full detailed analysis presented here. We assume that the
total ensemble of vortex lines is a superposition of both
contributions \cite{Jou-PRB74-2006}
\be\label{Pi-s-controErota}%
{\bf \Pi}^s=(1-b){\bf
\Pi}^s_H+b{\bf \Pi}^s_R , \hskip0.4in
 {\bf \Pi}^a= c{\bf \Pi}^a_R . \ee

In (\ref{Pi-s-controErota}),  $b$ and $c$ are parameters between 0
and 1, which depend on $\bf\Omega$ and ${\bf V}_{ns}$, describing
the relative weight of the array of vortex lines parallel to
$\bf\Omega$ and the disordered tangle of counterflow.

In Ref.~\cite{Jou-PRB74-2006},  we examined explicitly two
simplified situations: $\bf V$ parallel to $\bf\Omega$ and $\bf V$
orthogonal to $\bf\Omega$; in the first case there is cylindrical
symmetry with respect to the rotation axis; in the second one, no
symmetries are present in the vortex tangle.

In  the case ${\bf  V \|  \Omega}$, choosing $\bf V$ and
$\bf\Omega$ in the direction of the $x$ axis, one has for the
symmetric part of tensor $\bf\Pi$:
 \be\label{New-1}
{\bf\Pi}^s={\bf\Pi}_H^s+{\bf\Pi}_R^s= \frac{3}{2}(1-b)\left(
                \begin{array}{ccc}
                  2a& 0& 0  \\
                  0&1-a&0 \\
                  0&0&1-a  \\
                \end{array}
              \right)+\frac{3}{2}b\left(
                \begin{array}{ccc}
                  0& 0& 0  \\
                  0&1&0 \\
                  0&0&1 \\
                \end{array}
              \right)
 \ee
which can be written:
 \be\label{New-2}{\bf\Pi}^s={3\over 2}<{\bf
U}-{\bf s'\bf s'}> = {3\over 2}\left(
                \begin{array}{ccc}
                  2a(1-b)& 0& 0 \\
                  0&1-a(1-b)&0  \\
                  0&0&1-a(1-b)  \\
                \end{array}
              \right). \ee
Therefore, we have
 $<{s'_x}^2>=1-2a(1-b)$, $<{s'_y}^2> =
<{s'_z}^2>=a(1-b)$, and $I_\|=2a(1-b)$, $I_\bot=1- a(1-b)$. Once
known the coefficient $a$, from experiments in pure counterflow,
from (\ref{New-2}) we can obtain the dependence of $b$ from the
second moments of $\bf s'$.

 In the second situation, when the counterflow velocity
$\bf V$ is orthogonal to the angular velocity $\bf \Omega$,
choosing $\bf\Omega$ in the direction of $x$ axis and $\bf V$ in
the one of the $z$ axis, one has, for ${\bf\Pi}^s$:
 \be\label{New-3}{\bf \Pi}^s=(1-b){\bf \Pi}_H^s+b{\bf \Pi}_R^s=
{3\over 2} \left(\begin{array}{ccc}
                  (1-a )(1-b)& 0& 0   \\
                  0&(1-a )(1-b)+b&0  \\
                 0&0&2a (1-b)+b  \\
                \end{array}    \right)
 . \ee
  As a consequence, it results
\be\label{New-s} <{s'_x}^2>=a (1-b)+b , \hskip0.3in <{s'_y}^2> =a
(1-b) ,  \hskip0.3in <{s'_z}^2>=(1-2a )(1-b) . \ee

The antisymmetric part of tensor $\bf\Pi$. which depends only on
${\bf \Pi}_R$, furnishes the polarity of the tangle; in fact,
being $\bf W$ a constant tensor, one can write:
\be\label{New-4}{\bf\Pi}^a={3\over 2}{\alpha'\over\alpha} c
  {\bf W} \cdot{\bf \hat\Omega} = {3\over2}{\alpha'\over\alpha} c
\left(
  \begin{array}{ccc}
    0& 0& 0  \\
    0&0&1 \\
    0&-1&0 \\
  \end{array}
\right), \ee
 from which we deduce $< s'_x> ={\alpha'\over\alpha}c$, $<{s'_y}>=
<{s'_z}>=0$.

 A microscopic evaluation of coefficients $b$ and $c$ was made in
Ref.~\cite{Jou-PRB74-2006}, based on a paramagnetic analogy, which
reflects the competition between the orienting effects of
$\bf\Omega$ and the randomizing effects of ${\bf V}_{ns}$,
respectively analogous to the orienting effects of a magnetic field
${\bf H}$ on magnetic dipoles $\mu$ and the randomizing effects of
thermal excitations. In rotating counterflow, the rotation $\Omega$
orients the vortices along its direction, in an analogous way to
${\bf H}$, whereas the counterflow ${\bf V}_{ns}$ plays a
disordering role. We found, using the Langevin model of
paramagnetism,
\be\label{sprim-x2}%
<{s'_x}>=\coth x-{1\over x} ,\ee and
\be\label{sprim-xalla2}%
<{s'_x}^2>= 1+{2\over x}\left[ {1\over x}- \coth x \right],\ee
 with $x$ proportional to $ {\Omega\kappa/V_{ns}^2}$.
Similar situations would be found in Couette and Poiseuille flow
where the velocity gradient, instead that a rotation, contributes
to orient the vortices \cite{Swanson-PRL50-1983,Godfrey-PF13-2001}
with $\Omega$ replaced by the local shear rate.

A particular illustration of the combination of
(\ref{Pi-s-controErota}) and (\ref{Fns-new-finale}) could be
provided by the decay of small perturbations of ${\bf V}_n-{\bf
V}_s$. According to (\ref{equa-Vn}) and (\ref{equa-Vs}) and the
assumption ${\bf T=0}$ and $\nabla T\cong 0$, one would have, in a
linear approach
\be\label{Vn-Vs}%
{\partial({\bf V}_n-{\bf V}_s)\over \partial t }=-{2\over
3}\alpha{\rho \over \rho_n}\kappa L(1-b) ({\bf V}_n-{\bf
V}_s)+\frac{\eta}{\rho_n}\nabla^2{\bf V}_n , \ee being ${\bf
V}_n-{\bf V}_s$ along $x$ axis. The first term on the right hand
side, describes ${\rho\over \rho_s \rho_n} {\bf F}_{ns}$, taking
into account (\ref{Fns-new-finale}) and (\ref{Pi-s-controErota}).

 For long-wave perturbations, $\nabla^2{\bf V}_n$ will be small as
 compared to the first term and the decay of ${\bf V}_n-{\bf V}_s$
 to its steady state value will be exponential, with a relaxation
 time depending on the anisotropy parameter $b$. This time could
 provide a measurement of $b$ independent of the measurement
 provided by the attenuation of the second sound along different
 axis of the system. An analogous analysis could be carried out to
 explore the anisotropy parameter $a$ introduced in (\ref{Pi-H}) for
 tangles in pure counterflow.

A second particular illustration may underline the role of the
term in $\bf I$ in expression (\ref{Fns-new-2}) for ${\bf
F}_{ns}$.  This would be reflected in the temperature gradient
needed to maintain a steady state value of ${\bf V}_{ns}$, at
constant pressure. Under these conditions, one has
\be%
- 2\rho_s S  \nabla T +2 {\bf F}_{ns}+\eta\nabla^2 {\bf V}_n =0.
\ee
  Thus, neglecting, for simplicity, the term in $\eta\nabla^2 {\bf
V}_n$ and using (\ref{Fns-new-2}) for ${\bf F}_{ns}$, we have
\be%
\alpha \kappa L\left[ {2\over 3}{\bf \Pi} \cdot  {\bf
V}_{ns}+\tilde \beta c_1 L^{1/2} {\bf I}\right] =  S  \nabla T .
\ee
 It is clear that the presence of $\bf I$ modifies the relation
 between ${\bf V}_{ns}$ and $\nabla T$; this could allow to measure
the influence of the corresponding term.

These two simple illustrations show that the generalized equation
for  ${\bf F}_{ns}$ considered in (\ref{Fns-new-2}) is indeed
expected to have specific applications.

\section{Conclusions}
\setcounter{equation}{0} Summarizing, in this work we propose to
substitute in the HVBK equations the expression of the mutual
friction force ${\bf F }_{ns}$ (equation (\ref{Fns-HVBK})) with
the equation (\ref{Fns-new-finale}), to take into account of the
second-order moment of $\bf s'$, allowing in this way that not all
the vortex lines in the small volume element under consideration
have the same direction.  Analogously, the expression
(\ref{T-HALL}) for the tension $\bf T$ has been replaced by
equation (\ref{roT-new-finale}). The coefficients $b$ and $c$
appearing in (\ref{Pi-s-controErota}) can be related to the
counterflow velocity and to the angular velocity by the relations
(\ref{New-s}), with $<{s'_x}>$ and $<{s'_x}^2>$ expressed by
(\ref{sprim-x2}) and (\ref{sprim-xalla2}). For the vectors $\bf I$
and $\bf J$ we have chosen the constitutive relations (\ref{c1LI})
and (\ref{c1LJ}).

Another important modification  consists in adding to the
evolution equations for ${\bf V}_n$ and  ${\bf V}_s$ an evolution
equation for $L$  (see eq. (\ref{deL-su-dt3}), or
(\ref{deL-su-dt-Onsager2}) and (\ref{matrix-Onsager})), including
the effects of polarization anisotropy and inhomogeneities. In
fact, in a general situation, it is not correct in (\ref{kappa-L})
to substitute $\kappa L$ with the modulus of the curl of ${\bf
V}_s$. But, in a first approximation it should be valid when a
relevant polarization is present in the vortices, that is when the
polarity vector ${\bf p}$ approximate to  1. So, the results
obtained by some authors which measure the decay of $L$ or compute
a fluid viscosity by means of this relationship are good when the
polarization of the vortex tangle is high enough. Relations
(\ref{deL-su-dt3}) and (\ref{deL-su-dt-Nabla}) avoid this
simplification.

Our approach is not an exact description of superlfuid
hydrodynamics, but an approximation which, in the mentioned
situations, is more satisfactory that the HVBK assumption. In
fact, an exact description of the hydrodynamics would require not
only to analyze $\bf s'$ but also $\bf s''$, $\bf s'''$ and so on,
and to deal not only with their respective average values, but
also with their second and higher-order moments as well as with
their correlations. Something similar happens in classical
turbulence, in whose detailed description a hierarchy of equations
for the second and higher-order models of the fluctuations
velocity appears, and is usually truncated by some approximate
arguments. Here, instead of the moments of the random component of
the fluctuating velocity, the higher-order moments of $\bf s'$,
$\bf s''$, $\bf s'''$  and so on should appear, but this would be
not practical at all, and conveniently truncated approaches must
be analyzed.

The domain of validity of the new equations does not depend only
on $L$ but also on polarization, which may have different values
for a same value of $L$; for instance, in rotating counterflow, a
same $L$ may have different polarizations depending on the value
of the angular velocity; analogously, as mentioned above, towing
parallel thin cylinders or towing a grid of two orthogonal
families of parallel cylinders, but at different speeds, could
give a same $L$ but with polarization values different from each
other; finally, for short values of $L$, and not very developed
turbulence, pinned vortices may contribute to the polarization,
and could provide situations for an analysis.

A possible experiment of interest would be to compare the
turbulence produced by towing through the superfluid a set of very
thin cylinders parallel to each other, and that produced by towing
a grid composed of two orthogonal families of parallel thin
cylinders, perpendicular to each other. In the first case, a
strong polarization parallel to the cylinders is expected, whereas
for the grid much less polarization should be expected.
A fourth possible check would be, as mentioned in Section 4.1, the
influence of the tension $T$ in the dynamics of long wavelength
helical vortex lines.

In Section 5, we have concentrated our attention on rotating
counterflow; another illustration of the possible physical
consequences of the tension ${\bf T}$, may be found in steady
Poiseuille flows in an isothermal situation. There, addition of
(\ref{equa-Vn}) and (\ref{equa-Vs}) leads for the equation
describing the velocity profile of the normal component
\be\label{nabla}%
-\nabla p+\eta \nabla^2 {\bf V}_{n}+\rho_s {\bf
T}={\bf 0}.%
\ee Thus, if ${\bf T}={\bf 0}$, the ${\bf V}_n$ profile will have
the typical parabolic form of Newtonian fluid. Modifications of
fluid velocity profile have been studied by Godfrey and Barenghi
\cite{Godfrey-PF13-2001}. From (\ref{nabla}), these modifications
would be related to the form of ${\bf T}$. A completely polarized
array of parallel rectilinear vortices or a completely isotropic
vortex tangle would have ${\bf T}={\bf 0}$ and would not modify
the form of the ${\bf V}_n$ profile. However, net tension
different from zero could arise at relatively low fluxes, due to
the influence of vortices pinned to the walls, which in the
presence of the flow would have a curvature opposite  to the
velocity of the flow. In particular, they would have an influence
on a small-amplitude oscillating flow along a cylinder with pinned
vortices. An analysis of such situation would provide another
possible check of the equations proposed here. A third possible
check would be to study the decay of counterflow rotating
turbulence for different values of the angular speed, as it has
been mentioned in Section 5.

Finally, to mention yet another situation where the effects of the
polarization are important, and for which experimental data have
been recently available, is the eddy formation in two-dimensional
counterflow in the presence of a transverse cylinder, in which two
macroscopic vortices appear in the upstream part of the cylinder
\cite{Zhang-vanSciver}, in contrast with the usual vortices
appearing downstream of the cylinder in plane flows of normal
viscous fluids. These macroscopic vortices, rotating in opposite
directions, could arise from a separation of clockwise and
counterclockwise microscopic vortices induced by the presence of
the cylinder. Though this is, at present, only a plausible guess,
but not a definite explanation of this interesting phenomenon, it
allows to see, at least tentatively, a situation in which the
results of this work would be of special interest. Indeed, far
upstream from the cylinder, the flow would have vortices (i.e. $L$
different from 0), but null polarity. Near the cylinder, in the
macroscopic vortex regions, the microscopic vortices would be
thorouhgly polarized, in two opposite directions. In between,
there would be a process in which polarization would raise from 0
to 1. In the corresponding process, equation (2.4) for ${\bf
F}_{ns}$ cannot be used, because it does not account for any
change of polarization. Instead, in (4.4) the polarization is
present through the tensor $\Pi$. Explaining in full detail the
results of Ref \cite{Zhang-vanSciver} seems not straightforward,
but it seems a good field for application of the generalized
equations proposed in this paper.

\section*{Acknowledgments}
We acknowledge the support of the Acci\'{o}n Integrada
Espa\~{n}a-Italia (Grant S2800082F HI2004-0316 of the Spanish
Ministry of Science and Technology and grant IT2253 of the Italian
MIUR). DJ acknowledges the financial support from the
Direcci\'{o}n General de Investigaci\'{o}n of the Spanish Ministry
of Education under grant FIS 2009-13370-C02-01 and of the
Direcci\'{o} General de Recerca of the Generalitat of Catalonia,
under grant 2009 SGR-00164. MSM and MS acknowledge the financial
support by "Fondi 60\%" of the University of Palermo. MS
acknowledges the "Assegno di ricerca: Studio della turbolenza
superfluida e della sua evoluzione" of the University of Palermo.

\end{document}